\documentclass[superscriptaddress,secnumarabic,nobibnotes,aps,showpacs,showkeys,nofootinbib,preprint]{revtex4}

\usepackage[colorlinks=true, pdfstartview=FitV, linkcolor=blue, citecolor=red, urlcolor=magenta,breaklinks=true]{hyperref}
\usepackage{graphicx}
\usepackage{float}
\usepackage{epsf}
\usepackage{bm}
\usepackage{amsmath}
\usepackage{amsfonts}
\usepackage{amssymb}
\usepackage{epstopdf}
\usepackage{natbib}%
\usepackage{comment}
\newcommand{\pa}{\partial}

\newcommand{\bes}{\begin{subequations}}
\newcommand{\ees}{\end{subequations}}
\def\ben{\begin{eqnarray}}
\def\een{\end{eqnarray}}
\newcommand{\bens}{\begin{subeqnarray}}
\newcommand{\eens}{\end{subeqnarray}}

\def\be{\begin{equation}}
\def\ee{\end{equation}}

\def\exp{\text{exp}}

\def\ni{\noindent}

\begin{document}

\title{\Large Low-energy Chern-Simons-Proca theory} 
 
\author{Diego R. Granado}\email{diegorochagranado@duytan.edu.vn}
\affiliation{Institute of Research and Development, Duy T{a}n University, D{a} Nang 550000, Viet Nam}
\affiliation{Faculty of Natural Sciences,  Duy Tan University, Da Nang 550000, Viet Nam}

\author{Everton M. C. Abreu}\email{evertonabreu@ufrrj.br}
\affiliation{Departamento de F\'{i}sica, Universidade Federal Rural do Rio de Janeiro, 23890-971, Serop\'edica, RJ, Brazil}
\affiliation{Departamento de F\'{i}sica, Universidade Federal de Juiz de Fora, 36036-330, Juiz de Fora, MG, Brazil}
\affiliation{Programa de P\'os-Gradua\c{c}\~ao Interdisciplinar em F\'isica Aplicada, Instituto de F\'{i}sica, Universidade Federal do Rio de Janeiro, 21941-972, Rio de Janeiro, RJ, Brazil}

\author{A. Yu. Petrov}\email{petrov@fisica.ufpb.br}
\affiliation{Departamento de F\'{i}sica, Universidade Federal da Para\'{i}ba, Caixa Postal 5008, 58051-970, Jo\~{a}o Pessoa, Para\'{i}ba, Brazil}

%\author{Paulo J. Porfirio$^{a,b}$}
%\affiliation{$^{1}$ Institute of Research and Development, Duy T{a}n University, D{a} Nang 550000, Viet Nam}
%\affiliation{$^{2}$  Faculty of Natural Sciences,  Duy Tan University, Da Nang 550000, Viet Nam}
%\affiliation{$^{3}$Departamento de F\'isica, Universidade Federal de Juiz de Fora–UFJF, 36036-330, Juiz de Fora, MG, Brazil}
%\affiliation{$^{4}$Departamento de F\'isica, Universidade Federal Rural do, Rio de Janeiro–UFRRJ, 23890-971, Serop\'edica, RJ, Brazil}
%\affiliation{$^{5}$Programa de P\'os-Gradua\ c c\~ao Interdisciplinar em F\'isica Aplicada, Instituto de F\'isica, Universidade Federal do Rio de Janeiro-UFRJ, 21941-972, Rio de Janeiro, RJ, Brazil}
%\author{Antonio J. G. Carvalho$^{b}$}
%\author{A. Yu. Petrov$^{c}$}
%\author{Paulo J. Porfirio$^{a,b}$}
%%\affiliation{$^{a}$Department of Physics and Astronomy, University of Pennsylvania, Philadelphia, PA 19104, USA}
%\affiliation{$^{a}$ Institute of Research and Development, Duy T{a}n University, D{a} Nang 550000, Viet Nam}
%\affiliation{$^{b}$ Faculty of Information Technology,  Duy Tan University, Da Nang 550000, Viet Nam}

%\date{\today}

\begin{abstract}
\ni Some time ago, the infrared limit of the Abelian Chern-Simons-Proca theory was investigated. In this letter, we show how the Chern-Simons-Proca theory can emerge as an effective low energy theory. Our result is obtained by means of a procedure that takes into account the proliferation, or dilution, of topological defects presented in the system. %At the end of our letter we show also how an emergent massive higher-order derivative Chern-Simons theory appears.
\end{abstract}

\date{\today}

\maketitle

%%%%%%%%%%%%%%%%%%%%%%%

%%%%%%%%%%%%%%%%%%%%%%%%%%%%%%%%%%%
\section{Introduction}

In a very interesting paper, the infrared limit of the Abelian Chern-Simons-Proca (CSP) theory was investigated in \cite{Niemi:1994nw}. By starting with a regular Abelian CS theory and adding a Proca mass term to the gauge field, the consequence was the introduction of a new dimensional parameter which is an important perturbation of the CS action at the infrared, or long-distance, limit. As it was pointed in \cite{Niemi:1994nw}, this procedure could have relevant consequences in the quantum Hall effect or in high-temperature superconductivity, where it could, for instance, parametrize finite size effects in an experimental framework. Also in \cite{Niemi:1994nw}, it was found that in the infrared limit of the Abelian CSP theory, the topological quantum mechanical models
investigated in \cite{Dunne:1989hv} can be recovered.

In this letter, we have used the Julia-Toulouse (JT) mechanism to achieve an action whose low energy degrees of freedom are determined by the Abelian CSP theory. %Indeed, it is well known \cite{CW} that the JT mechanism is based on the coupling of the gauge field to some extra fields, the topological defects, represented by some sources $J^{\mu}$. 
The idea behind the JT mechanism is the description of how the dynamics of the topological defects presented in the system can influence it. 
Superconductors can be seen as one example of this procedure. Proliferation of vortices on the sample can destroy the superconductivity state. As it will be formally presented in the next section, JT mechanism can be used to describe this phase transition between a superconductivity state and Maxwell free theory due to vortices proliferation. In order to achieve this result, an activation term, which looks like $\eta^{\mu}{\cal O}\eta_{\mu}$, and describing the vortices dynamics, is introduced.  Here, $\eta$ is the field responsible for introducing vortices in the system, so that the integration over $\eta$ brings new terms to the gauge theory. 

In \cite{CW,CWours,Braga:2016atp}, this operator ${\cal O}$ was treated as a derivative expansion. In these papers, the goal was to study the influence of topological defects in the lower regime of the theory. As such, the authors only considered the low energy fluctuations of $\eta$. As a consequence a Thirring-like interaction term was generated. Here we also took into account the low energy fluctuations but we considered a Chern-Simons (CS) interaction term for the currents. From this point, we demonstrated that, as a result, we have a theory whose lower energy regime is described by the Abelian CSP theory. 

%a proliferation of the topological defects in a system, such that they become {dynamical} fields, i.e., condensation, drives the {system} to a phase transition. superconductors can be seen as a good example to describe this idea. In the case of a superconductor, a proliferation of defects can drive the system from a free Maxwell theory to a gauge theory describing the system in a superconductivity state. 

This paper is organized as follows: in section \ref{secI}, we reviewed the work presented in \cite{Braga:2016atp} about the condensation of topological defects in regular superconductors.    In section \ref{emergentnonlocalterm}, we show how the mechanism used to describe superconductors can be used to obtain an electrodynamics whose low energy regime is described by the Abelian CSP theory.   Finally, in the section \ref{sec-com}, we present our comments and conclusions.

%==============================================================================
\section{Short review on the condensation of topological defects}
\label{secI}
%==============================================================================

%In this section, we review the mechanism presented in \cite{Braga:2016atp} {to explain} how a condensation of topological defects can drive a system to a phase transition.
In this section, we present a short review on how the JT mechanism describes a superconductor phase transition\footnote{A detailed review can be found in \cite{Braga:2016atp,Granado:2019bqk}}.  As well known, the proliferation of vortices in a superconductor sample can destroy the superconductivity state. Hence, the system goes from a superconductivity state to a Maxwell free system. In our approach, the proliferation of vortices, such that it destroys the superconductivity state, we call it as vortices condensation.   Otherwise, we name it vortices dilution. In summary, in the dilution scenario, the system stays as a perfect superconductor and in condensation scenario, free Maxwell theory is recovery. The JT mechanism describes the dilution/condensation process. The following partition function 
%\begin{widetext}
 \begin{align}
 \label{dilutecharges3dual}
 Z[j] =\sum_K \int {\cal D} A \int {\cal D} \theta\;  e^{-\int d^4x \left( \frac{1}{4} F_{\mu\nu}F^{\mu\nu} + \frac{q^2m^2}{2}\left(  A_{\mu} + \frac{1}{q} \partial_{\mu}\theta+\frac{2\pi}{q}K_\mu\right)^2 \right)} e^{-ie\int d^4x \; A_{\mu}j^{\mu}  } 
\end{align}
%\end{widetext}
describes the dilution/condensation process, where $K$ stands for the vortices. Condensation can be described as the case where $K$ becomes a continuous field. An integration over $K$ can be computed and the result is a free Maxwell theory will be obtained. The partition function in Eq. \eqref{dilutecharges3dual} can be derived from
%\begin{widetext}
 \begin{align}
 \label{dilutecharges3dual2}
 Z[j] =\sum_K \int {\cal D} A \int {\cal D} \theta\int {\cal D} \eta\;  e^{-\int d^4x \left( \frac{1}{4} F_{\mu\nu}F^{\mu\nu} - iq\left(  A_{\mu} + \frac{1}{q} \partial_{\mu}\theta+\frac{2\pi}{q}K_\mu\right)\eta \right)} e^{-ie\int d^4x \; A_{\mu}j_{\mu}}e^{-S_\eta} \,\,. 
\end{align}
%\end{widetext}
%
From the simplest example, where $S_\eta=0$ in Eq. \eqref{dilutecharges3dual2}, the reason why $K$ represents a vortice is clear. In this case, $\eta$ is a Lagrange multiplier. Thus the gauge field is restricted to flux filaments.  The presence of $S_\eta$ in Eq. \eqref{dilutecharges3dual2} is connected to both the  contact terms and the topological dynamics of the system \cite{Braga:2016atp}. The action for $\eta$ in $S_\eta$, in Eq. \eqref{dilutecharges3dual2}, to obtain Eq. \eqref{dilutecharges3dual}, is such that 
\begin{align}
 \label{Seta1}
  S_{\eta} = \frac{1}{2m^2}\eta_{\mu}\eta^{\mu} + \frac{1}{2m^2 m^2}\eta_{\mu}\Box \eta^{\mu} + \cdots \,\,,
\end{align}
where $m$ is a massive parameter.  We will show later that $m$ is linked with the penetration length of the superconductor. In order to obtain Eq. \eqref{dilutecharges3dual}, only the first term needs to be considered. As it was pointed out in \cite{Braga:2016atp}, the expansion in Eq. \eqref{Seta1} is considered because the goal was to describe the low energy fluctuations of the field $\eta$. The action in Eq. \eqref{dilutecharges3dual} is the one who describes a superconductor with penetration length $1/m$. Thus, as it was stated before, $m$ is linked to the penetration length of the superconductor sample. 
\section{Low-energy effective Chern-Simons-Proca theory}
\label{emergentnonlocalterm}
%==============================================================================

%Lorentz symmetry breaking is a high energy phenomenon, meaning that even though it is possible to create a low energy effective action who presents such a feature, the emergence of such a term can only be obtained by considering high energy effects. The constant vector defining the theory preferred direction can be introduced in the high energy fluctuations of the system. Thus in the low energy regime the theory is Lorentz invariant. In our scenario the emergent low (high) energy electromagnetic effective action is determined by the integration over $\eta$ by considering only the first term in \eqref{Seta1}. 

From the previous section, it can be seen that if we consider the low energy fluctuations of $\eta_\mu$, we are able to describe the electromagnetic response in a superconductor. The goal of this section is to use the formalism described before to show how an effective CSP theory can be obtained in the low energy regime. 

In \cite{Granado:2019bqk, Braga:2016atp} the authors explored the derivative expansion concerning Eq. \eqref{Seta1}. 
In \cite{Granado:2019bqk}, it was shown that {for the case of} high energy fluctuations {of} $\eta_\mu$, i.e., the action for $\eta_\mu$ in Eq. \eqref{Seta1} can be written as  $S_{\eta}=\eta_\mu\sum_n (\Box/m^2)^n\eta^\mu$. In this paper, we dealt with the lower energies fluctuations of $\eta_\mu$, but now we will consider at least one derivative. This can be described in the following manner
%From \eqref{dilutecharges3dual}, we know in advance that new system will emerge only in the condensate phase. 
\begin{eqnarray}
 \label{Seta2}
  S_{\eta}&=&\frac{1}{2m^2} \eta_{\mu}\eta^{\mu} + \frac{\xi}{2m^2} \eta_{\mu}(\epsilon^{\mu\nu\rho}\partial_\rho) \eta_{\nu} + \cdots \,\,,
%&=&  \frac{1}{2m^2} \eta_{\mu}\left[\sum_{n=0}^\infty\xi^n\left(a.\partial\right)^n\right]\eta_{\mu}
%&=&  \frac{1}{2m^2} \eta_{\mu}{\cal F}\left({\Box}/{m^{2}_p}\right)\eta_{\mu}
\end{eqnarray}
%\begin{equation}
%\eta_\nu=\frac{-iqm^2}{(1-{\xi}^22\partial^2)}\left(\delta_{\sigma\delta}-\frac{\partial_\sigma\partial_\delta}{\partial^2}-\xi\epsilon_{\sigma\delta\rho}\partial_\rho\right)A_\sigma
%\end{equation}
%
where $\xi$ is a massive parameter. As it will be shown later, this parameter is linked to the coupling constant for the CS term. From the dimensional analysis, $\xi$ is proportional to $1/m$ and $\xi$ can be connected to a superconductor penetration length. In fact, all the massive parameters in Eq. \eqref{Seta2} could be rewritten in terms of $\xi$. In comparison to Eq. \eqref{Seta1}, the new term represents a CS dynamics for the $\eta$ field. This possibility was considered before in \cite{Braga:2016atp} for the case of different $\eta$'s fields. By integrating out $\eta_\mu$, as a result we have
%\begin{widetext}
\begin{eqnarray}
S &=& \int d^3x \Bigg( \frac{1}{4} F_{\mu\nu}F^{\mu\nu} +\frac{q^2m^2}{(1-2{\xi}^2\partial^2)}A'_{\mu}A'^{\mu}+ \frac{2qm^2\xi^2}{(1-2{\xi}^2\partial^2)}A'_{\mu}{\partial^\nu\partial^\mu}A'_\nu \nonumber \\
&+&\frac{qm^2}{(1-2{\xi}^2\partial^2)}\xi\epsilon^{\mu\nu\rho}A'_\nu\partial_\rho A'_\mu \Bigg)\,\,.
\label{pqed}
\end{eqnarray}
%\end{widetext}
%\begin{widetext}
%\begin{align}
%S = \int d^3x \left( \frac{1}{4} F_{\mu\nu}F_{\mu\nu} -\frac{q^2m^2}{(1-{\xi}^22\partial^2)}A_{\mu}A_{\mu}+ \frac{qm^2}{(1-{\xi}^22\partial^2)}A_{\mu}\frac{\partial_\nu\partial_\mu}{\partial^2}A_\nu+\frac{qm^2}{(1-{\xi}^22\partial^2)}\xi\epsilon_{\sigma\delta\rho}A_\delta\partial_\rho A_\sigma \right)
%\label{pqed}
%\end{align}
%\end{widetext}
%
By means of the change of variable, explained in Appendix \ref{appendixa}, the action in Eq. \eqref{pqed}, using the substitution of ${\cal H}\left({\Box}\right)^{-1/2}A'_\mu\to\tilde{A}_{\mu}$ becomes
\begin{widetext}
 \begin{align}
\label{themodel1}
S=\int d^3x \left(\frac{1}{4} F_{\mu\nu}{\cal H}\left({\Box}\right)F^{\mu\nu}+ \frac{2qm^2\xi^2}{2}A'_{\mu}{\partial^\nu\partial^\mu}A'_\nu
+\theta\epsilon^{\mu\nu\rho}A'_\nu\partial_\rho A'_\mu + \frac{q^2m^2}{2}A'_\mu A'^\mu \right)\,\,,%+\frac{1}{2\alpha}\left(\partial_\mu 
\end{align}
\end{widetext}
%\begin{widetext}
%\begin{align}
%S = \int d^3x \left( \frac{1}{4} F_{\mu\nu}F_{\mu\nu} -\frac{q^2m^2}{(1-{\xi}^22\partial^2)}A_{\mu}A_{\mu}+ \frac{qm^2}{(1-{\xi}^22\partial^2)}A_{\mu}\frac{\partial_\nu\partial_\mu}{\partial^2}A_\nu+\frac{qm^2}{(1-{\xi}^22\partial^2)}\xi\epsilon_{\sigma\delta\rho}A_\delta\partial_\rho A_\sigma \right)
%\label{pqed}
%\end{align}
%\end{widetext}
%
where $\theta$ is defined by $\theta=qm^2\xi$. Thus, in the Lorentz gauge, as a result we have a massive Maxwell-Podolsky-CS theory. The CS term here resemble an axion term. A study involving this action will be addressed in a future work. The goal in this letter is to investigate how the CSP theory can emerge as an effective low energy theory. As it is known, as higher is the order of the derivative, more relevant is the term in the ultraviolet regime of the theory. Thus, in the low energy regime of the theory, the most relevant terms are the ones who has one, or no derivatives. Therefore, in the Lorentz gauge $\partial^\mu A_\mu=0$, from Eq. \eqref{themodel1}, we can see that the most relevant terms in low energy limit are %the %effective action ones, namely,
%\begin{widetext}
 \begin{align}
\label{themodel2}
S_{eff}=\int d^3x \left(\theta\epsilon^{\mu\nu\rho}A'_\nu\partial_\rho A'_\mu + \frac{q^2m^2}{2}A'_\mu A'^\mu \right) \,\,,%+\frac{1}{2\alpha}\left(\partial_\mu 
\end{align}
%\end{widetext}
%
%This redefinition was done in order to make the notation more clear. 
The action in Eq. \eqref{themodel2} is precisely the one in \cite{Niemi:1994nw}. In \cite{Niemi:1994nw}, it was shown that Eq. \eqref{themodel2} is precisely the topological CS quantum mechanical model analyzed in \cite{djt}. 

\section{Comments and conclusions}
\label{sec-com}
%==============================================================================
 
Through the last decades, the Abelian and non-Abelian versions of $D=3$ CS theory have found several applications in several areas of physics.   The gauge invariant modification of the Abelian CS theory, discussed here, obtained by adding a mass term to the gauge field was analyzed in \cite{Niemi:1994nw}. 

  %Due to our approach, the mass term added in \cite{Niemi:1994nw} can be linked to the penetration length of a regular superconductor. A physical interpretation to such a massive parameter can be find here.   
In this letter, we have used the JT approach to analyze a system whose low energy degrees of freedom are described by the Abelian CSP theory. In our approach, the CSP theory lives in the infrared regime of a massive Maxwell-Podolsky-CS theory. In \cite{Granado:2019bqk}, the authors use the JT mechanism to obtain a massive Maxwell-Podolsky theory. In this paper it was pointed out that, in the case of a massive Maxwell-Podolsky theory, the massive parameter $m$ is still connected to the superconductor penetration length. Here we obtain a massive Maxwell-Podolsky-CS theory. Due to the presence of the CS term a careful analysis needs to be addressed and this will be presented in a future work. 

The procedure of introducing a massive Proca term brought together a new dimensionfull parameter, which is an interesting perturbation of the CS action in the infrared limit. As it was pointed in \cite{Niemi:1994nw}, there could be underlying consequences, for example, in the quantum Hall effect or high-temperature superconductivity. As it was stated before, the JT mechanism was used to describe regular superconductors and topological superconductors in \cite{Braga:2016atp}. As perspective of future research papers, we intend to investigate the role of the CSP in high-temperature superconductivity phenomena. This opens up a new road for the application the JT mechanism in the superconductivity matters. 

\appendix
\section{Change of variables and Integration over $\eta$}\label{appendixa}
%%%%%%%%%%%%%%%%%%%%%%%%%%%%%%%%%%%%%%%%%%%%%%%%%%%%%%%%%%%%%%%%%%

In this Appendix we will review the change of variables introduced in \cite{Granado:2019bqk}. By means of a change of variables, the nonlocality included in the potential term can be transferred to the kinetic term and vice-versa. For instance, let us consider the following Lagrangian
\begin{eqnarray}
{\cal L}&=&\frac{1}{2}\pa_m\phi\, \exp\,\Bigg(\frac{\Box}{2\mu^2}\Bigg)\pa^m\phi-V(\phi)\,\,.
\end{eqnarray}
As it can be seen, the nonlocality is presented in the kinetic term. Let us do the change of variables
\begin{eqnarray}
\exp \Bigg(\frac{\Box}{2\mu^2}\Bigg) \,\phi\,\to\,\tilde{\phi}\,\,.
\end{eqnarray}
It is clear that the Jacobian of this change of variable is $e^{\frac{\Box}{2\mu^2}}\delta(x-y)|_{x=y}$, which is a constant whose presence modifies  trivially the generating functional. Besides, we assume that this change of variables is invertible, i.e., it is performed only with the fields defined in such a manner that a non-zero field, namely, a field which is not equal identically to zero, is mapped to a non-zero field only. It implies that if $\phi|_{x=y}=0$, i.e., $\phi$ vanishes at some surfaces only, we have $e^{\frac{\Box}{2\mu^2}}\phi|_{x=y}=0$, and the same is true for higher derivatives of $\phi$. In other words, if the field or some linear combination of fields and its derivatives vanishes at some surface, its image also will vanish on the same surface.
Our Lagrangian takes the form
\begin{eqnarray}
{\cal L}&=&\frac{1}{2}\pa_m\tilde{\phi}\pa^m\tilde{\phi}-V\Big( e^{-\frac{\Box}{2\mu^2}}\,\tilde{\phi}\Big).
\end{eqnarray}
So, the potential becomes nonlocal instead of the kinetic term, just like in certain cases when the kinetic term looks like $\phi\Box\widehat{T}\phi$, and the potential term looks like $((\widehat{T})^{1/2}\phi)^n$, where $\widehat{T}$ is an operator introducing the nonlocality. %f.e. $\widehat{T}=e^{\frac{\Box}{2\mu^2}}$ as in the example above, we can remove the nonlocality both from kinetic and potential %term, but these cases are trivial). This can be generelized for other field theory models, including the gauge field.

%We note that within all these our replacements by the rule $\phi\Box\widehat{L}\phi$, with $\widehat{L}$ is the nonlocal operator, and the same rule for $A_{\mu}$, no extra contributions to the effective actions are generated. Indeed, when we carry out these transformations,  although there are nonlocal, they are linear in fields, so, in the generating functional we get only the extra multiplier $\det\hat{L}^{1/2}$, and since it does not depend on any fields, it yields only a field-independent additive term in the effective action which clearly can be neglected. 

%%%%%%%%%%%%%%%%%%%%%%%%%%%%%%%%%%%%%%%%%%%%%%%%%%%%%%%%%%%%%%%%%%%%%%%
%\section{Integration over $\eta$}\label{appendixb}
%%%%%%%%%%%%%%%%%%%%%%%%%%%%%%%%%%%%%%%%%%%%%%%%%%%%%%%%%%%%%%%%%%%%%%%
Considering both terms in Eq. \eqref{Seta2}, we can solve the path integral for $\eta$ by using its equation of motion. The equation of motion for $\eta$ reads
%\begin{eqnarray*}
%&&\frac{\delta}{\delta\eta_\sigma} \left(-iqA_\mu\eta_\mu-\frac{1}{2m^2} \eta_{\mu}\eta^{\mu} - \frac{\xi}{2m^2} \eta_{\mu}(\epsilon^{\mu\nu\rho}\partial_\rho) \eta_{\nu}\right)=\\
%&
%=
%&iqA_\sigma+\frac{1i}{2m^2}\eta_\sigma+\frac{\xi}{2m^2} \frac{\delta}{\delta\eta_\sigma} \eta_{\mu}(\epsilon^{\mu\nu\rho}\partial_\rho) \eta_{\nu}\\
%&
%=
%&iqA_\sigma+\frac{1}{2m^2}\eta_\sigma+\frac{\xi}{2m^2} \left(\epsilon^{\sigma\nu\rho}\partial_\rho \eta_{\nu}-\epsilon^{\mu\sigma\rho}\partial_\rho \eta_{\mu}\right)\\
%&
%=
%&iqA_\sigma+\frac{1}{m^2}\eta_\sigma+\frac{\xi}{m^2} \epsilon^{\sigma\nu\rho}\partial_\rho \eta_{\nu}\\
%\end{eqnarray*}
%The equations of motions:
\begin{eqnarray}
%\left(\delta_{\sigma\nu}+{\xi} \epsilon^{\sigma\nu\rho}\partial_\rho\right)\eta_\nu&=&-iqm^2A_\sigma\nonumber\\
B_{\sigma\nu}\eta_\nu&=&-iqm^2A'_\sigma
\label{eom}
\end{eqnarray}
where
\begin{equation}
\label{opb}
B_{\sigma\nu}=\left(\delta_{\sigma\nu}+{\xi} \epsilon^{\sigma\nu\rho}\partial_\rho\right)
\end{equation}
In order to find the solution of Eq. \eqref{eom}, we have to compute the inverse operator of Eq. \eqref{opb}. This operator is not  invertible. In this way we have to supplement Eq. \eqref{Seta2} such that
%\begin{eqnarray*}
%  S_{\eta}&=&\frac{1}{2m^2} \eta_{\mu}\eta^{\mu} + \frac{\xi}{2m^2} \eta_{\mu}(\epsilon^{\mu\nu\rho}\partial_\rho) \eta_{\nu}+ \frac{1}{2m^2}(\partial_\mu \eta_\mu)^2 + \cdots\\
%&=&\frac{1}{2m^2} \eta_{\mu}\eta^{\mu} + \frac{\xi}{2m^2} \eta_{\mu}(\epsilon^{\mu\nu\rho}\partial_\rho) \eta_{\nu}+ \frac{1}{2m^2}(\partial_\mu \eta_\mu)(\partial_\nu \eta_\nu) + \cdots\\
%&=&\frac{1}{2m^2} \eta_{\mu}\eta^{\mu} + \frac{\xi}{2m^2} \eta_{\mu}(\epsilon^{\mu\nu\rho}\partial_\rho) \eta_{\nu}-\frac{1}{2m^2}\eta_\mu \partial_\mu \partial_\nu \eta_\nu + \cdots
%\end{eqnarray*}
%where from \eqref{pathint-maxwell} we have that $\partial_\mu\eta_\mu=0$. Now the equations of motion reads:
%\begin{eqnarray*}
%\eta_\sigma+{\xi} \epsilon^{\sigma\nu\rho}\partial_\rho \eta_{\nu}-\partial_\sigma \partial_\nu \eta_\nu&=&-iqm^2A_\sigma\\
%\left(\delta_{\nu\sigma}+{\xi} \epsilon^{\sigma\nu\rho}\partial_\rho -\partial_\sigma \partial_\nu\right) \eta_\nu&=&-iqm^2A_\sigma\\
%B_{\nu\sigma}\eta_\nu&=&-iqm^2A_\sigma
%\end{eqnarray*}
%where
\begin{equation*}
B_{\nu\sigma}=\left(\delta_{\nu\sigma}+{\xi} \epsilon^{\sigma\nu\rho}\partial_\rho +\left(1/\alpha-1\right)\partial_\sigma \partial_\nu\right) \,\,,
\end{equation*}
and now $B_{\nu\sigma}$ is invertible. In order to show this we have that 
%\begin{equation*}
$B_{\nu\sigma}B_{\sigma\delta}^{-1}=\delta_{\nu\delta}$.
%\end{equation*}
Let us construct an ansatz
\begin{equation}
B_{\sigma\delta}^{-1}=A\delta_{\sigma\delta}+B\partial_\sigma\partial_\delta+C\epsilon_{\sigma\delta\rho}\partial_\rho \,\,.
\label{inverse}
\end{equation}
%\begin{widetext}
%\begin{eqnarray*}
%B_{\nu\sigma}B_{\sigma\delta}^{-1}&=&\delta_{\nu\delta}A+B\partial_\nu\partial_\delta+C\epsilon_{\nu\delta\rho}\partial_\rho\\
%&
%+
%&
%{\xi} \epsilon^{\delta\nu\rho}\partial_\rho A+{\xi}C \epsilon^{\sigma\nu\alpha}\epsilon_{\sigma\delta\rho}\partial_\alpha\partial_\rho+A\left(1/\alpha-1\right)\partial_\delta \partial_\nu+B\left(1/\alpha-1\right) \partial^2 \partial_\nu\partial_\delta\\
%&
%=
%&
%\delta_{\nu\delta}A+B\partial_\nu\partial_\delta+C\epsilon_{\nu\delta\rho}\partial_\rho\\
%&
%+
%&
%{\xi} \epsilon^{\delta\nu\rho}\partial_\rho A+{\xi}2C(\delta_{\nu\delta}\delta{\alpha\rho}-\delta{\nu\rho}\delta{\alpha\delta})\partial_\alpha\partial_\rho\\
%&
%+
%&
%A\left(1/\alpha-1\right)\partial_\delta \partial_\nu+B\left(1/\alpha-1\right) \partial^2 \partial_\nu\partial_\delta\\
%&=&\delta_{\nu\delta}A+B\partial_\nu\partial_\delta+C\epsilon_{\nu\delta\rho}\partial_\rho\\
%&+&{\xi} \epsilon^{\delta\nu\rho}\partial_\rho A+{\xi}2C\delta_{\nu\delta}\partial^2-{\xi}2C\partial_\nu\partial_\delta\\
%&+&A\left(1/\alpha-1\right)\partial_\delta \partial_\nu+B\left(1/\alpha-1\right) \partial^2 \partial_\nu\partial_\delta=\delta_{\nu\delta}
%\end{eqnarray*}
%\end{widetext}
From this we have the following set of equations
%\begin{eqnarray*}
%A+{\xi}2C\partial^2&=&1\\
%B(1+\left(1/\alpha-1\right) \partial^2)&=&{\xi}2C-A\left(1/\alpha-1\right)\\
%C+A\xi&=&0\\
%\end{eqnarray*}
\begin{eqnarray*}
A(1-2{\xi}^2\partial^2)&=&1\\
B&=&\frac{{\xi}2C-A\left(1/\alpha-1\right)}{(1+\left(1/\alpha-1\right) \partial^2)}\\
C&=&-A\xi \,\,.
\end{eqnarray*}
%\begin{eqnarray*}
%A+{\xi}2C\partial^2&=&1\\
%B-{\xi}2C+A\left(1/\alpha-1\right)+B\left(1/\alpha-1\right) \partial^2&=&0\\
%C+A\xi&=&0\\
%\end{eqnarray*}
%From these relations we have
%\begin{widetext}
%\begin{eqnarray*}
%A&=&\frac{1}{(1-{\xi}^22\partial^2)}\\
%\alpha B-\alpha{\xi}2C+A\left(1-\alpha\right)+B\left(1-\alpha\right) \partial^2&=&0\\
%\alpha B-\alpha{\xi}2C+A-A\alpha+B\partial^2-B\alpha\partial^2&=&0\\
%B&=&-\frac{A}{\partial^2}-\alpha (B-{\xi}2C-B\partial^2-A)\\
%C&=&-A\xi\\
%\end{eqnarray*}
%\end{widetext}
For $\alpha=1$, we have that Eq. \eqref{inverse} reads
\begin{equation}
B_{\sigma\delta}^{-1}=\frac{1}{(1-{\xi}^22\partial^2)}\left(\delta_{\sigma\delta}-2\xi^2{\partial_\sigma\partial_\delta}-\xi\epsilon_{\sigma\delta\rho}\partial_\rho\right) \,\,.
\end{equation}
%\begin{equation}
%B_{\sigma\delta}^{-1}=\frac{1}{(1-{\xi}^22\partial^2)}\left(\delta_{\sigma\delta}-\frac{\partial_\sigma\partial_\delta}{\partial^2}-\xi\epsilon_{\sigma\delta\rho}\partial_\rho\right)
%\end{equation}
The solution of Eq. \eqref{eom} reads
\begin{equation}
\eta_\nu=\frac{-iqm^2}{(1-{\xi}^22\partial^2)}\left(\delta_{\sigma\delta}-2\xi^2{\partial_\sigma\partial_\delta}-\xi\epsilon_{\sigma\delta\rho}\partial_\rho\right)A'_\sigma \,\,.
\end{equation}
%

%\begin{widetext}
% \begin{align}
%\label{themodel1}
%S=\int d^3x \left(\frac{1}{4} F_{\mu\nu}{\cal H}\left({\Box}\right)F^{\mu\nu}+ \frac{\lambda}{2}A'_{\mu}\frac{\partial_\nu\partial_\mu}{\partial^2}A'_\nu+\theta\epsilon_{\sigma\delta\rho}A'_\delta\partial_\rho A'_\sigma + \frac{q^2m^2}{2}A'_\mu A'_\mu \right).%+\frac{1}{2\alpha}\left(\partial_\mu 
%\end{align}
%\end{widetext}
%So, we generated the essentially nonlocal Maxwell term.

\section*{Acknowledgments}

The authors thank CNPq (Conselho Nacional de Desenvolvimento Cient\' ifico e Tecnol\'ogico), Brazilian scientific support federal agency, for partial financial support, grants numbers  301562/2019-9 (A.Yu.P.) and 406894/2018-3 (E.M.C.A.).

\end{document}